# Superconductivity in doped FeTe$_{1-x}$S$_x$ (x= 0.00 to 0.25) single crystals

P. K Maheshwari[1,2], V. Raghavendra Reddy[3], B. Gahtori[1] and V.P.S. Awana[1,2,*,#]

[1]CSIR-National Physical Laboratory, Dr. K. S. Krishnan Marg, New Delhi-110012, India
[2] AcSIR- Academy of Scientific & Innovative Research-NPL, New Delhi-110012, India
[3]UGC-DAE Consortium for Scientific Research, Khandwa Road, Indore-425001

**Abstract**

We report self flux growth and characterization of FeTe$_{1-x}$S$_x$ (x= 0.00 to 0.25) single crystal series. Surface X-ray diffraction (XRD) exhibited crystalline nature with growth in (00$l$) plane. Micro-structural (electron microscopy) images of representative crystals showed the slab-like morphology and near stoichiometric composition. Powder XRD analysis (Rietveld) of single crystals exhibited tetragonal structure with P4/nmm space group and decreasing $a$ and $c$ lattice parameters with increase in x. Electrical resistivity measurements ($\rho$-$T$) showed superconductivity with $T_c^{onset}$ at 9.5K and 8.5K for x =0.10 and x =0.25 respectively. The un-doped crystal, i.e., x =0.00, exhibited known step like anomaly at around 70K. Upper critical field $H_{c2}(0)$, as calculated from magneto transport [$\rho(T)H$], for x =0.25 crystal is around 60Tesla and 45Tesla in $H//ab$ and $H//c$ directions. Thermal activation energy [$U_0(H)$] calculated for x =0.10 and 0.25 crystals followed weak power law, indicating single vortex pinning at low fields. Mossbauer spectra for FeTe$_{1-x}$S$_x$ crystals at 300K and 5K are compared with non superconducting FeTe. Both quadrupole splitting ($QS$) and isomer shift ($IS$) for S doped crystals were found to decrease. Also at 5K the hyperfine field for x =0.10 superconducting crystal is decreased substantially from 10.6Tesla (FeTe) to 7.2Tesla. For x =0.25 crystal, though small quantity of un-reacted Fe is visible at room temperature, but unlike x =0.10, the low temperature (5K) ordered FeTe hyperfine field is nearly zero.

*__Corresponding Author__
Dr. V. P. S. Awana:  E-mail:awana@nplindia.org
Ph. +91-11-45609357, Fax-+91-11-45609310 Homepage:awanavps.webs.com
#ORCID; https://orcid.org/0000-0002-4908-8600

**Introduction**

Fe based compounds were mainly known for their magnetic properties and observation of superconductivity in them was thought to be a far cry. However, the scenario changed dramatically after the discovery of superconductivity in Iron based compounds, first in Iron pnictides [1-2] and later in Iron chalcogenides [3]. This attracted huge attention of scientific community in the field of superconductivity. Iron based superconducting compounds and high $T_c$ [*HTSc*] Cuprates [4-5] are known to be outside the conventional BCS (Bardeen Cooper & Schrieffer) theory of superconductivity [6]. Another superconductor discovered after the high $T_c$ Cuprates, i.e., $MgB_2$ yet follows the BCS in its strong coupling limit [7].

The Fe chalcogenide compounds are known to possess most simple crystal structure among other Fe based superconductors. The ground state of Fe chalcogenides i.e., FeTe orders anti-ferromagnetically along with coupled structural/magnetic phase transitions at below around 70K. The coupled structural/magnetic phase transition of FeTe is seen clearly in electrical, thermal and magnetic measurements at around the same temperature [8-10]. The substitution of S/Se at Te site in FeTe introduces superconductivity by suppressing the anti-ferromagnetic ordering [8, 11-14]. The highest superconducting temperature ($T_c$) for Fe(Te/Se, S) chalcogenides series is achieved at approximately 15K for $FeSe_{0.50}Te_{0.50}$ at ambient conditions [8, 15]. Further, the $T_c$ of Fe chalcogenide systems, increases up to 37K under high pressure [16] and to above 50K with favourable alkali metal intercalations [17-19].

The single crystal growth of Fe chalcogenide superconductors is rather difficult and mainly the self flux Bridgman and the added flux (KCl/NaCl) with complicated heat treatments are employed for crystal growth [20-24]. Bridgman method is quite expensive as the same employs state of art furnaces. On the other hand flux method is marred with foreign contamination, which is difficult to remove completely; hence self flux method is preferred to grow quality single crystals. As far as growth of $FeTe_{1-x}S_x$ single crystals is concerned the same is rather difficult due to large difference between the ionic radius of $Te^{-2}$(211pm) and $S^{-2}$ (186pm) [25]. Here, we report self flux crystal growth and superconductivity characterization of $FeTe_{1-x}S_x$ (x= 0.00 to 0.25) series. After, repeated runs, we are been able to grow the single crystals of $FeTe_{1-x}S_x$ series by self flux method. Superconductivity occurs in $FeTe_{1-x}S_x$ with increase in Te site S substitution and the magnetic ordering (~70K) for FeTe gets gradually suppressed. Host of physical properties including structural details, morphology, superconductivity and Mossbauer spectroscopy are reported here for $FeTe_{1-x}S_x$ single crystal series. In our view, this is detailed study of large single crystal of $FeTe_{1-x}S_x$ series using self flux method. Earlier only tiny/flux assisted single crystals or poly-crystals of FeTe$_{1-x}$S$_x$ series were reported [12, 14, 25]. Further, worth mentioning is the fact that although Fe chalcogenide superconductivity is heavily studied for both poly and single crystal of FeTe$_{1-x}$Se$_x$ [3, 8-10, 15], the same is only scant for FeTe$_{1-x}$S$_x$ [12, 14, 25].

**Experimental Details**

All the studied FeTe$_{1-x}$S$_x$ (x = 0.0 to 0.25) single crystals are grown via melt process in a simple programmable furnace using self flux method from constituent elements of 99.99 purity. Well mixed stoichiometric powders were pelletized, vacuum sealed in quartz tube and kept in a furnace. The heating schedule involved various steps at $1000^0$C, $200^0$C, $500^0$C and $750^0$C in case of S doped crystals [26] and slightly simpler route for pristine x =0.0 crystal [27]. Ref. 26 deals with only x =0.10 Sulfur doped crystal and very few results obtained at that time on the same were reported in a conference. The same heat treatment as used for x =0.10 is applied for x =0.05 and 0.25, which resulted in good crystals. Beyond x =0.25 we could not grow the S doped FeTe$_{1-x}$S$_x$ crystals. The reason may be large difference between the ionic radius of Te$^{-2}$(211pm) and S$^{-2}$ (186pm) [25]. Basically, the crystals are grown from melt of mixed and well pulverized stoichiometric constituent elements by slow cooling with good range of Sulfur solubility at Te site of up to 25%. The transport and magnetization results are obtained on Quantum Design (QD) Physical Property Measurement System (PPMS) down to 2K under applied magnetic field of up to 14 Tesla. $^{57}$Fe Mossbauer measurements were done on crushed powders of the synthesized single crystals in transmission mode having $^{57}$Co as radioactive source.

**Results and Discussion**

Heat treatment schedule employed for single crystal growth of FeTe$_{1-x}$S$_x$ (x = 0.05 to 0.25) is shown by the schematic diagram in Figure 1. This heat treatment is different than the one being employed for growth of FeTe or FeTe$_{1-x}$Se$_x$ single crystals (shown in right inset of Fig. 1). Only, after several trials, we could optimize the final heating schedule to grow large (cm size, left inset of Fig. 1) FeTe$_{1-x}$S$_x$ (x = 0.0 to 0.25) single crystals. The heat treatment thus employed for FeTe$_{1-x}$S$_x$ was checked for reproducibility and found alright.

SEM (Scanning electron microscopy) images and EDX (Energy dispersive X-ray analyzer) are useful for understanding the morphology and elemental analysis of the synthesized samples. SEM and EDX had been performed at room temperature for FeTe$_{0.75}$S$_{0.25}$ single crystal and the results are shown in Figure 2. It is clear from the SEM image that FeTe$_{0.75}$S$_{0.25}$ crystallizes in layered structure, which is similar to the case of FeTe [27]. The EDX analysis is shown in insets (a) and (b) of Fig.2. Inset (a) shows the quantitative elemental analysis of selected area for $FeTe_{0.75}S_{0.25}$, which is found to be near to the stoichiometric ratio i.e., close to $FeTe_{0.75}S_{0.25}$. Inset view (b) depicts selected area spectral analysis of $FeTe_{0.75}S_{0.25}$, showing that all the elements of $FeTe_{0.75}S_{0.25}$ are being present without contamination of any other foreign element in the matrix.

Figure 3 depicts the room temperature XRD pattern carried out on the surface of $FeTe_{1-x}S_x$ (x=0.0 to 0.25) single crystals, indicating the crystal growth of all the samples in (00*l*) plane only. This result confirms the evidence of single crystalline property of the as grown $FeTe_{1-x}S_x$ (x=0.0 to 0.25) series. Further to study the detailed structural properties, we performed the powdered XRD followed by Rietveld refinement using Fullprof software. The observed powder XRD of all the studied $FeTe_{1-x}S_x$ crystals is shown in Fig. 4(a). All the samples of $FeTe_{1-x}S_x$ (x =0.0 to 0.25) series exhibit tetragonal structure within P4/nmm space group. Also seen were small impurities of crystalline $FeTe_2$. The values of lattice parameters, co-ordinate positions, and volume obtained from Rietveld refinement are depicted in Table 1. As the S concentration at Te site of $FeTe_{1-x}S_x$ (x =0.0 to 0.25) series increases, a monotonic decrease in lattice parameters (*a* and *c*) is clearly observed, which is in confirmation with our previous results on bulk polycrystalline samples of the same series [28]. The lattice parameters, *a* and *c* are seen to decrease from 3.82Å to 3.79Å and 6.29Å to 6.21Å for x =0.00 and x =0.25 respectively. Inset of Fig 4(a) shows the zoomed part of powder XRD of crystal series from $2\Theta$ angle of $42^0$ to $46^0$, the result clearly shows shifting of peak towards the higher angle for the [003] plane, representing decrement in the *c* lattice parameter with S doping.

Fig 4(b) depicts the Rietveld refinement of $FeTe_{1-x}S_x$ (x =0.0 to 0.25) series at room temperature. Here $\chi^2$ represents the goodness of fitting of the observed and experimental data, which is in single digits. The values of $\chi^2$ for respective samples are shown in figure 4(b) as well. Although Rietveld refinement of FeTe is reported earlier [27], but the same is shown here as well for the sake of inter comparison with others samples. From the Rietveld refinement, we can safely conclude that the entire essential phase is identified and there is only small impurity ($FeTe_2$) being marked by * is found in some of the crystals; particularly in x = 0.05 and 0.10. Further small impurity of $Fe_3O_4$ is found in x = 0.25 crystal, which is marked by #.

Figure 5 despites the resistivity ($\rho$) verses temperature (*T*) measurement for $FeTe_{1-x}S_x$ (x =0.0 to 0.25) crystals in temperature range from room temperature (300K) down to 2K and the $\rho$ is normalized at 300K i.e. $\rho(T) / \rho_{300}$. Fig. 5 clearly shows a semiconducting to metallic behavior for all the crystals except x =0.25, which showed only semiconducting to superconducting transition. The x =0.05 and x =0.10 crystals also showed superconducting transition at low temperatures. For pure FeTe crystal without applying any external field, a coupled magnetic/structural transition appears at around 70K and 65K during warming and cooling cycles respectively. Warming/cooling cycle hysteresis of transition width ($\Delta T$) of ~5K in temperature occurs during measurement due to the existence of latent heat throughout the cycles, which is in confirmation with our previous report [27, 29].

For FeTe crystal, neither $T_c^{onset}$ nor $T_c^{offset}$ ($\rho=0$) were seen down to 2K and rather a metallic step is seen below 70K due to magnetic transition [27]. For x =0.05, i.e. FeTe$_{0.95}$S$_{0.05}$ a semiconducting to metallic step occurs in $\rho$-$T$ measurement approximately at around 55K, followed by $T_c^{onset}$ at approximately 9K, but no $T_c^{offset}$ ($\rho=0$) down to 2K. In case of x=0.10 i.e., FeTe$_{0.90}$S$_{0.10}$ crystal $T_c^{onset}$ is 9.5K and $T_c^{offset}$ ($\rho=0$) is found to be 6.5K along with the metallic step due to magnetic ordering at around 48K. For x=0.25, i.e., FeTe$_{0.75}$S$_{0.25}$, the normal state metallic part associated with magnetic ordering is not seen and rather a semiconducting to superconducting transition occurs at $T_c^{onset}$ of around 8.5K and $T_c^{offset}$ ($\rho=0$) at 4.5K. Inset of Fig. 5 depicts the zoomed view of $\rho$-$T$ plots, showing $T_c^{onset}$, $T_c^{offset}$ and magnetic transition of studied FeTe$_{1-x}$S$_x$ (x=0.0 to 0.25) single crystals. It is clear from the $\rho(T)$ results that though FeTe shows only the metallic step due to reported magnetic ordering at 70K, the x=0.05 and 0.10 both exhibit the same but along with superconductivity onset at lower temperatures. This means in case of x =0.05 and 0.10 both magnetic ordering and superconductivity co-exist. Interestingly, In case of the x =0.25 the magnetic ordering associated metallic step is absent and the crystal is superconducting below 10K.

Further to understand the superconducting response of FeTe$_{1-x}$S$_x$ single crystals, $\rho$-$T$ measurements under applied magnetic field have been performed for FeTe$_{0.75}$S$_{0.25}$ single crystal under applied magnetic field of up to 14Tesla in both direction i.e., $H//ab$ and $H//c$ and the results are shown in Fig. 6(a) and 6(b). The broadening of the $H//c$ is wider in comparison to $H//ab$ field direction, this shows that superconductivity is relatively weak in $c$-direction. The anisotropy of superconductivity being visible in $H//ab$ and $H//c$ measurements further confirm the single crystalline nature i.e., uni-directional growth of the studied material.

The upper critical field $H_{c2}(0)$ for $H//ab$ and $H//c$ direction is calculated for FeTe$_{0.75}$S$_{0.25}$ single crystal. At absolute zero temperature, $H_{c2}(0)$ is determined by using Ginzburg Landau (GL) equation i.e., $H_{c2}(T) = H_{c2}(0)[(1 - (T/T_c)^2)/(1 + (T/T_c)^2)]$. Thus calculated $H_{c2}(0)$ of FeTe$_{0.75}$S$_{0.25}$ single crystal for $H//ab$ and $H//c$ directions are shown in inset view of Fig. 6(a) and 6(b) respectively. Further, $H_{c2}(0)$ is calculated using normal state resistivity criteria ($\rho_n$) of 10%, 50% and 90%, which is found to be around 25Tesla, 45Tesla and 60Tesla respectively for *H//ab* direction and 20Tesla, 35Tesla and 45Tesla for *H//c* direction for FeTe$_{0.75}$S$_{0.25}$ single crystal. The fitted solid lines imply the extrapolated curve fitting of the GL equation, i.e. $H_{c2}(T) = H_{c2}(0)[(1 - t^2)/(1 + t^2)]$, here t $=T/T_c$ is called reduced temperature and $T_c$ is superconducting transition temperature. From fitted curve, the resultant $H_{c2}(0)$ value is far exceeded from the Pauli Paramagnetic limit i.e. 1.84 times of $T_c$ [30]. High *$H_{c2}(0)$* of FeTe$_{0.75}$S$_{0.25}$ single crystal indicates the highly pinned superconductor against the external applied magnetic field.

For determination of the coherence length $\xi(0)$ of FeTe$_{0.75}$S$_{0.25}$ single crystal, we use the relation of $\xi(0)$ and $H_{c2}(0)$ as follows; $H_{c2}(0)=\varphi_0/2\Pi\xi(0)^2$, here $\varphi_0$ is flux quantum, i.e., $2.0678 \times 10^{-15}$Tesla-m$^2$. For FeTe$_{0.75}$S$_{0.25}$ single crystal, at absolute zero temperature, the calculated coherence length $\xi(0)$ from above equation comes to be 23.4Å and 27.05Å for *H//ab* and *H//c* directions.

For further analysis of the $\rho(T)H$ behaviour of the FeTe$_{1-x}$S$_x$ (x =0.0 to 0.25) series, the thermally activated flux flow (TAFF) plots, i.e., Ln$\rho$ verses $1/T$ at various applied fields are drawn for FeTe$_{0.90}$S$_{0.10}$ and FeTe$_{0.75}$S$_{0.25}$ single crystals and are shown in Fig. 7(a) and Fig. 7(b) respectively. Inset of Fig 7(a) shows the $\rho(T)H$ measurement of FeTe$_{0.90}$S$_{0.10}$ sample, which is reported earlier [26]. According to TAFF theory [31, 32], the TAFF region is described with the help of Arrhenius relation [33] i.e., Ln$\rho(T,H)$ = Ln$\rho_0(H)$ - $U_0(H)/k_BT$, here Ln$\rho_0(H)$ is temperature dependent constant, $U_0(H)$ activation energy and k$_B$ the Boltzmann constant. Ln$\rho$ vs $1/T$ graph would be linearly fitted in TAFF region for both the samples as described with the above equation. TAFF fitted plots of up to 12Tesla in case of FeTe$_{0.90}$S$_{0.10}$ and up to 14Tesla in case of FeTe$_{0.75}$S$_{0.25}$ are shown in Fig. 7(a) and Fig. 7(b) respectively. All the linearly fitted extrapolated lines with magnetic fields are intercepted at the same temperature, which coincides approximately at superconducting transition temperature ($T_c$) of the compound. The resistivity broadening pattern under magnetic field in both samples is quite similar to FeSe$_{0.50}$Te$_{0.50}$ superconductor [34] and is due to the thermally assisted flux motion [35].

The thermal Activation energy [$U_0(H)$] calculated for FeTe$_{0.90}$S$_{0.10}$ and FeTe$_{0.75}$S$_{0.25}$ single crystals for different magnetic fields is shown in Fig. 9(b). The thermal activation energy varies from 14meV to 5.2meV for magnetic field range from 0.5Tesla to 12Tesla for FeTe$_{0.90}$S$_{0.10}$ and 3.5meV to 1.33meV for magnetic field from 1Tesla to 14Tesla for FeTe$_{0.90}$S$_{0.25}$ single crystal. The calculated thermal activation energy is far lesser than the activation energy for FeSe$_{0.50}$Te$_{0.50}$ single crystal [34]. Thermal activation energy follows as power law relation i.e., $U_0(H) = K \times H^{-\alpha}$, here $U_0$ is thermal activation energy, $K$ is constant, $H$ is magnetic field and $\alpha$ is called field dependent constant. For lower magnetic field i.e., up to 2Tesla, $\alpha$ comes around 0.07 while for higher magnetic field i.e., greater than 4Tesla, $\alpha$ comes around 0.63 for FeTe$_{0.90}$S$_{0.10}$ single crystal. While for FeTe$_{0.75}$S$_{0.25}$ sample, $\alpha$ comes 0.12 up to 3Tesla magnetic field and 0.75 for high magnetic field of above 4Tesla. This result of weak power law dependence of $U_0(H)$ shows that single vortex pinning is effective at low fields for studied FeTe$_{0.90}$S$_{0.10}$ and FeTe$_{0.75}$S$_{0.25}$ single crystals [32,36].

In order to even further explore the superconducting properties of FeTe$_{1-x}$S$_x$, the magnetization measurements have been done down to 2K for FeTe$_{0.90}$S$_{0.10}$ single crystal, which are shown in Fig.8. The real ($M'$) and imaginary ($M''$) parts of AC susceptibility for FeTe$_{0.90}$S$_{0.10}$ single crystal are taken at 10Oe amplitude and 333Hz frequency cooled down to 2K in the absence of any DC magnetic field. From the figure, it is clearly seen that a sharp decrement occurs in real part of AC susceptibility below its $T_c$, i.e., at around 7K, showing the diamagnetic shielding of the sample, confirming the bulk superconductivity of the sample. Further below $T_c$, a peak occurs in imaginary part of AC susceptibility i.e. M'', that's reflecting the flux penetration effect inside the crystal. There is also no indication of two peak behaviors, thus ruling out the granularity or polycrystalline nature of studied crystal. Inset of Fig.8 shows the isothermal $MH$ plot of the FeTe$_{0.90}$S$_{0.10}$ single crystal at 2K, exhibiting a clear opening of loop at up to 1Tesla. This $MH$ plot shows the evidence of typical type-II superconductivity in our studied FeTe$_{0.90}$S$_{0.10}$ single crystal.

The Mossbauer spectra of FeTe$_{1-x}$S$_x$ (x =0.00 to 0.25) have been taken at 300K (RT) and 5K, which are shown in Fig 9. The detailed Mossbauer analysis of x=0.00, i.e. FeTe at 5K and RT is reported elsewhere [37], and is included here for sake of inter comparison. The room temperature (RT) Mossbauer spectra of x =0.05 and x =0.10 samples show an asymmetric paramagnetic doublet representing the existence of two Fe sites, which is similar to FeTe Mossbauer spectra at room temperature [37, 38]. The RT data is fitted with a doublet and a singlet for x =0.05 and x =0.10 samples. At RT after fitting the curve, values of quadrupole splitting ($QS$) and isomer shift ($IS$) are found to be in the range of 0.40 ± 0.02mm/s and 0.39 ± 0.01mm/s respectively for majority doublet and $IS$ for singlet is found to be 0.35 ± 0.03mm/s for both x =0.05 and x =0.10 sample. The $IS$ decreases monotonically with doping from 0.46 ± 0.01mm/s for FeTe [37] to 0.35 ± 0.03mm/s for FeTe$_{0.90}$S$_{0.10}$ sample. The fraction of the singlet is found to increase in x =0.10 sample (~ 40%) as compared to x =0.05 (~10%). However, for x =0.25 sample in addition to the asymmetric doublet, a magnetically split sextet is observed with a hyperfine field of about 32.8 ± 0.1Tesla, which could be due to presence of Fe$_3$O$_4$ [39]. Right frames of Fig. 9 show the Mossbauer spectra of all the samples measured at 5K. For x =0.05 and x =0.10 samples, in addition to the asymmetric doublet distribution of hyperfine fields resulting in a broad magnetic sextet is observed. Therefore, the data is fitted with distribution of hyperfine fields and also considering the paramagnetic doublet. An average hyperfine field of about 9.4Tesla and 7.2Tesla is observed from the fitting and the fraction of magnetic sextet is observed to be about 55% and 64% for x =0.05 and x =0.10 samples respectively. For x =0.25 sample, it is observed that the line width of central doublet increased at 5K as compared to 300K. This can be interpreted as due to the presence of small internal hyperfine fields in x =0.25 sample at 5K. It may be noted here that in un-doped FeTe complete magnetic splitting with hyperfine field of 10.6Tesla is seen [37], which decrease to 9.4Tesla and 7.2Tesla for x =0.05 and x =0.10. Therefore, with S doping in FeTe the magnetic ordering seems to be suppressed as observed from 5K Mossbauer data.

**Conclusion**

We have successfully synthesized FeTe$_{1-x}$S$_x$ (x=0.0 to 0.25) single crystals series by self flux method i.e., without any added flux. Host of structural (XRD), micro-structural (SEM), superconducting i.e., high field low temperature down to 2K transport and magnetization along with spectroscopy (Mossbauer) studies are carried out and reported here. It is seen that FeTe$_{1-x}$S$_x$ superconductors are very robust against magnetic field as their superconductivity is hardly affected by the same. Mossbauer spectroscopy results concluded that the ground state compound (FeTe) magnetic order gets disappeared and superconductivity sets in with doping of S at Te site.

**Acknowledgement**

Authors would like to thank their Director NPL India for his keen interest in the present work. P. K. Maheshwari thanks CSIR, India for research fellowship and AcSIR-NPL for Ph.D. registration.

**Table 1:** FeTe$_{1-x}$S$_x$ (x= 0.00 to 0.25) single crystals lattice parameters and coordinate positions using Rietveld refinement.

|  | **x= 0.00** | **x=0.05** | **x=0.10** | **x=0.25** |
|---|---|---|---|---|
| **a=b (Å)** | 3.826(2) | 3.803(2) | 3.801(2) | 3.798(2) |
| **c (Å)** | 6.292(3) | 6.242(3) | 6.237(3) | 6.212(3) |
| **V(Å$^3$)** | 92.1325(3) | 90.304(3) | 90.06(2) | 89.6405(2) |
| **Fe** | (3/4,1/4,0) | (3/4,1/4,0) | (3/4,1/4,0) | (3/4,1/4,0) |
| **Te** | (1/4,1/4,0.286) | (1/4,1/4,0.289) | (1/4,1/4,0.288) | (1/4,1/4,0.280) |
| **S** | - | (1/4,1/4,0.289) | (1/4,1/4,0.288) | (1/4,1/4,0.280) |

## FIGURE CAPTIONS

**Figure 1:** Schematic heat treatment diagram to grow of FeTe$_{1-x}$S$_x$ (x=0.05 to 0.25) single crystal series. Left Inset view is picture of synthesized crystal; right inset view is heat treatment of FeTe single crystal.

**Figure 2:** Selected SEM image of FeTe$_{0.75}$S$_{0.25}$ single crystal at room temperature. Inset view is EDX result of FeTe$_{0.75}$S$_{0.25}$ single crystal (a) Quantitative analysis and (b) Elemental spectral analysis.

**Figure 3:** Single crystal XRD for FeTe$_{1-x}$S$_x$(x=0.0 to 0.25) series at room temperature.

**Figure 4:** (a) Powder XRD patterns for crushed powders of FeTe$_{1-x}$S$_x$(x=0.0 to 0.25) series at room temperature. Inset view is zoomed [003] plane view for the same. (b) Observed and fitted XRD of FeTe$_{1-x}$S$_x$(x=0.0 to 0.25) series at room temperature using Rietveld refinement.

**Figure 5:** Normalized electrical resistivity ($\rho/\rho_{300}$) versus temperatureplots for FeTe$_{1-x}$S$_x$ (x=0.0 to 0.25) series in temperature range of 2K to 300K. Inset view is zoomed part of same in temperature range of 2K to 80K.

**Figure 6:** Temperature dependent electrical resistivity $\rho(T)$ under various magnetic field (a) FeTe$_{0.75}$S$_{0.25}$ single crystal for $H//ab$ direction. Inset view is calculated $H_{c2}$ in $H//ab$ direction for FeTe$_{0.75}$S$_{0.25}$ single crystal (b) FeTe$_{0.75}$S$_{0.25}$ single crystal for $H//c$ direction. Inset is $H_{c2}$ in $H//c$ direction for FeTe$_{0.75}$S$_{0.25}$ single crystal.

**Figure 7:** (a) $\ln\rho(T,H)$ vs $1/T$ for different magnetic fields for FeTe$_{0.90}$S$_{0.10}$ single crystal corresponding fitted solid line of Arrhenius relation. (b) Thermally Activation energy $U_o(H)$ with solid lines fitting of $U_o(H) \sim H^\alpha$ for different magnetic field for FeTe$_{0.90}$S$_{0.10}$ single crystal.

**Figure 8:** AC magnetic susceptibility of FeTe$_{0.90}$S$_{0.10}$ single crystal at 333Hz frequency and 10Oe amplitude from temperature range 10K to 2K. Inset view is MH plot of same at 2K.

**Figure 9:** Mossbauer spectra of FeTe$_{1-x}$S$_x$(x=0.00 to 0.25) series. Left frame at 300K (RT) and right frame at 5K temperature.

Fig 1

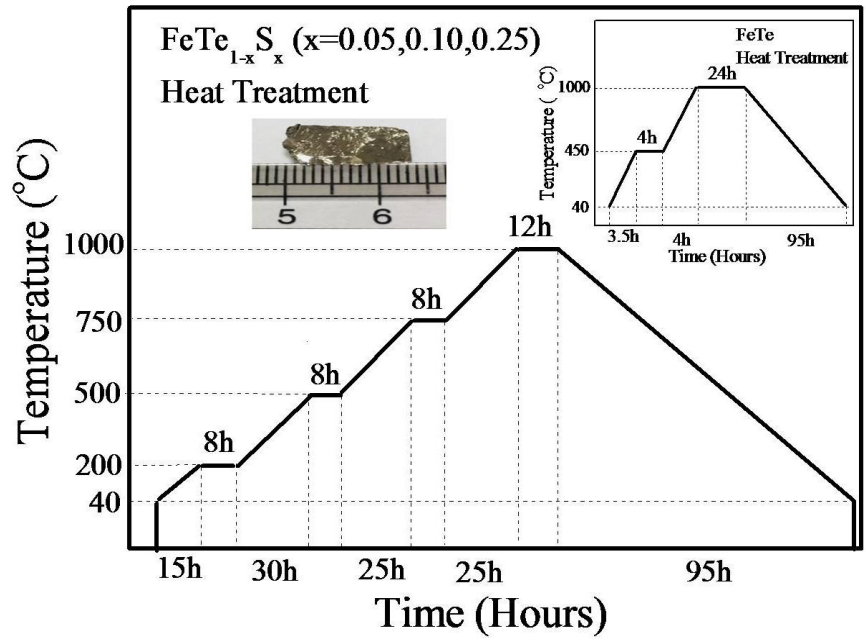

Fig 2

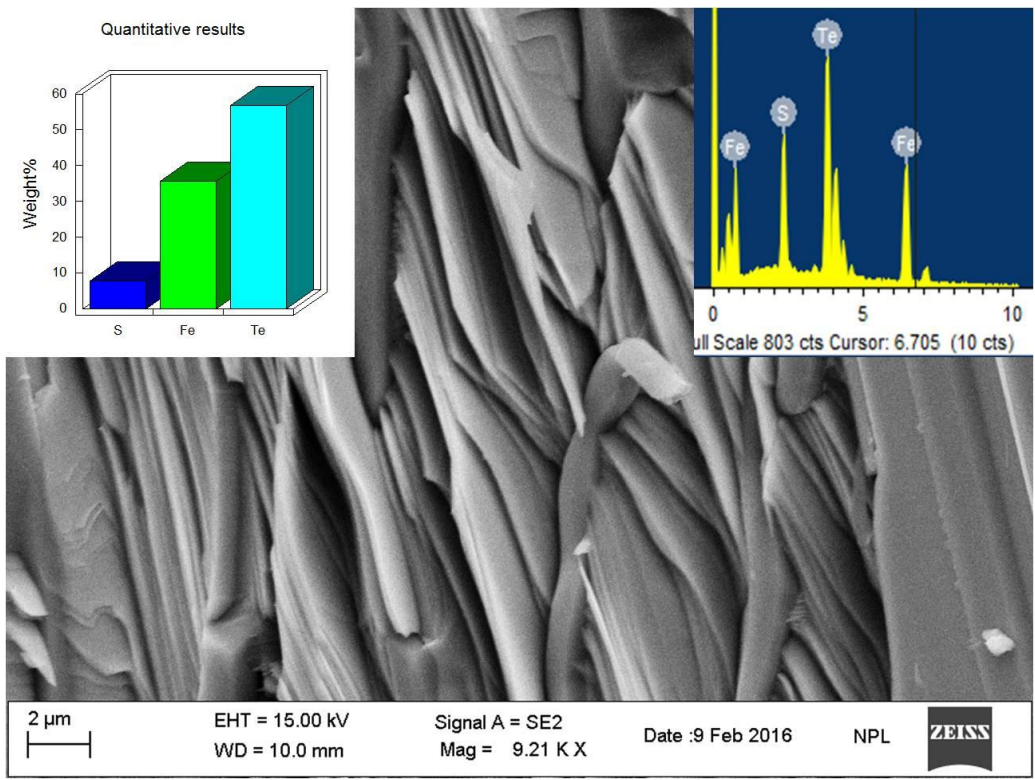

Fig 3

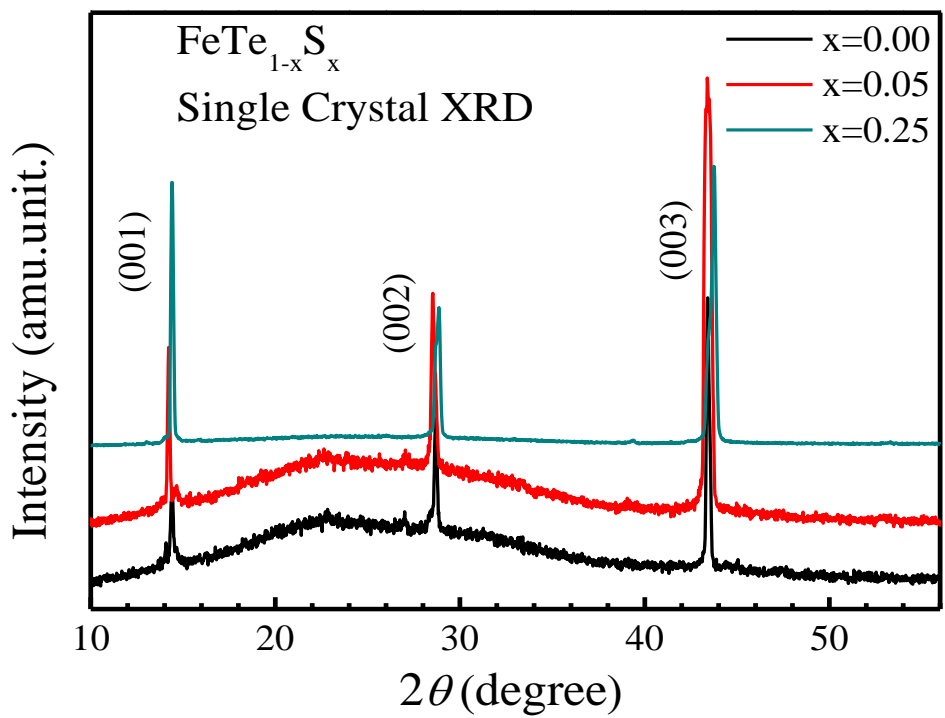

Fig 4(a)

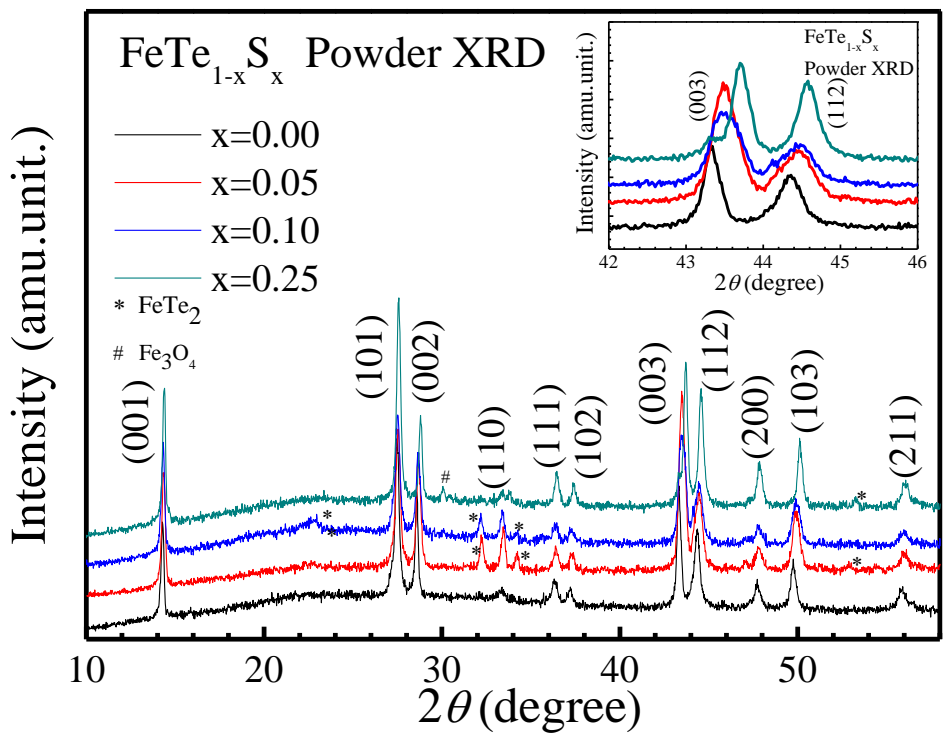

Fig 4(b)

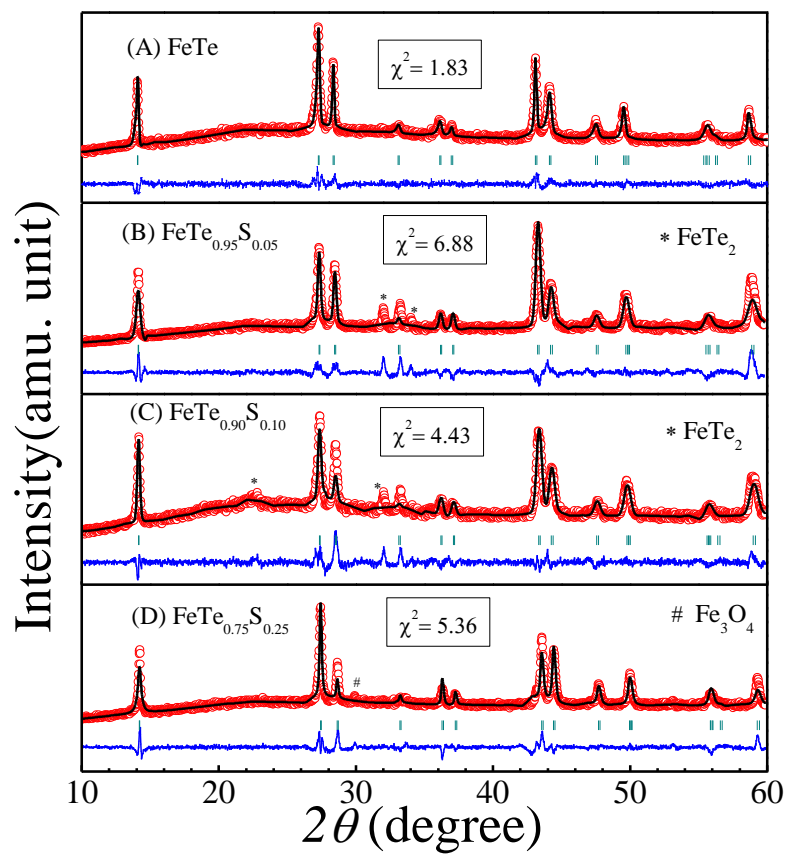

Fig 5

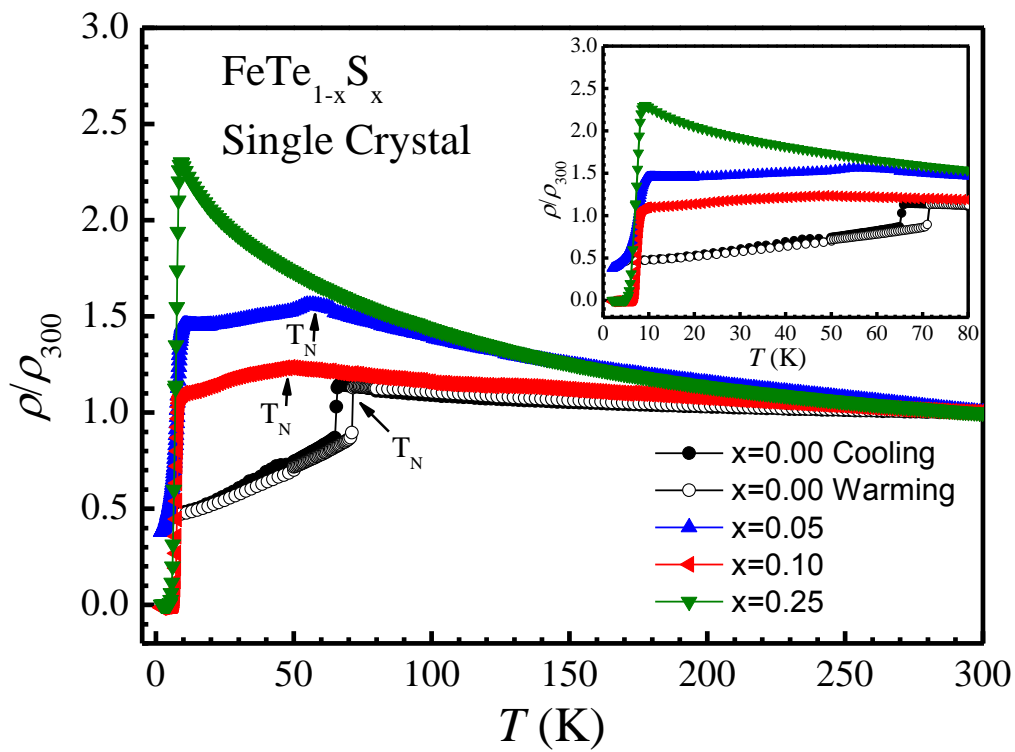

Fig 6(a)

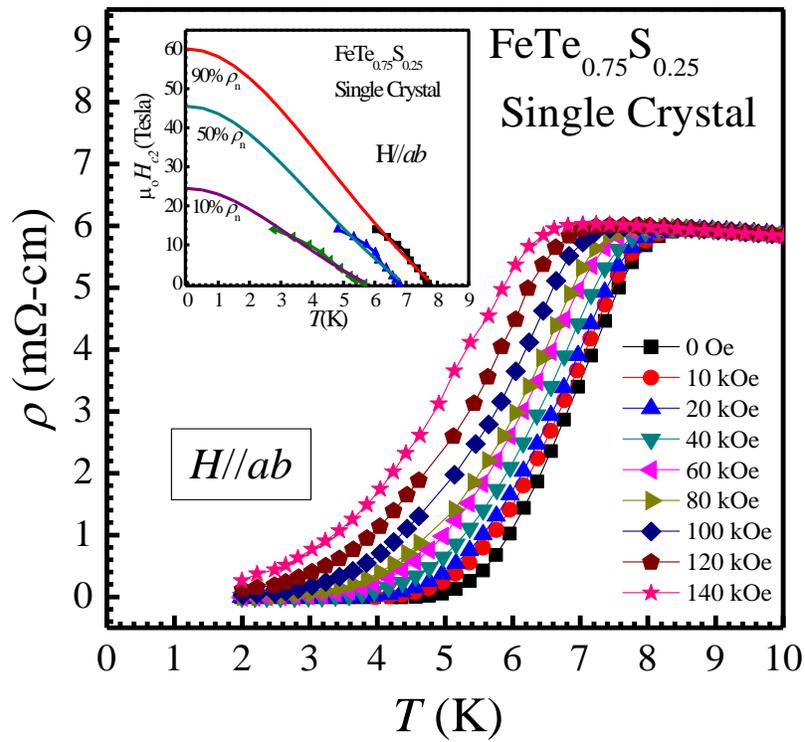

Fig 6(b)

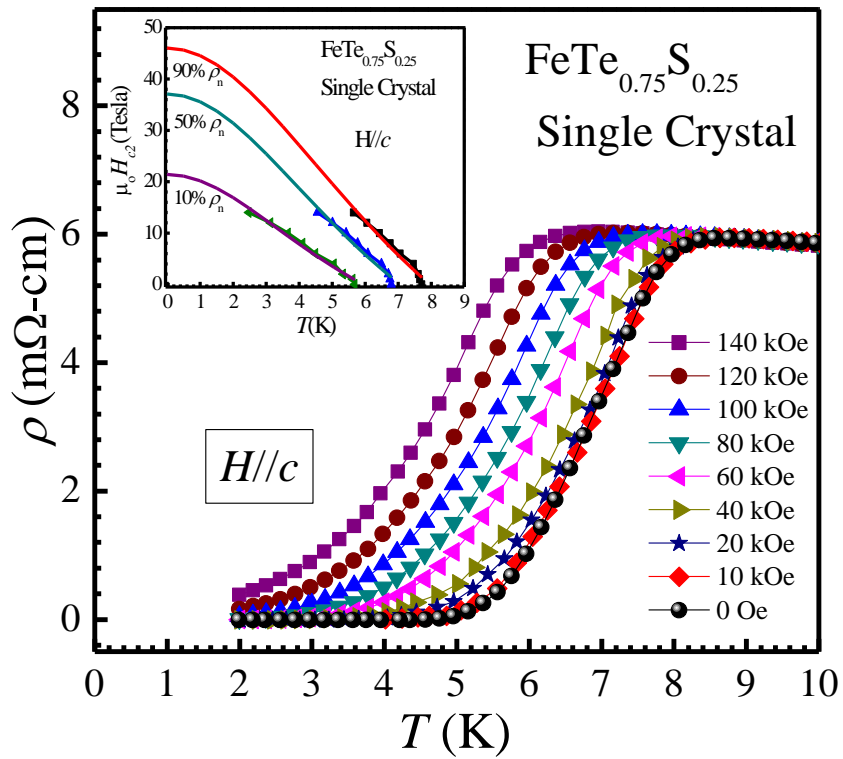

Fig 7(a)

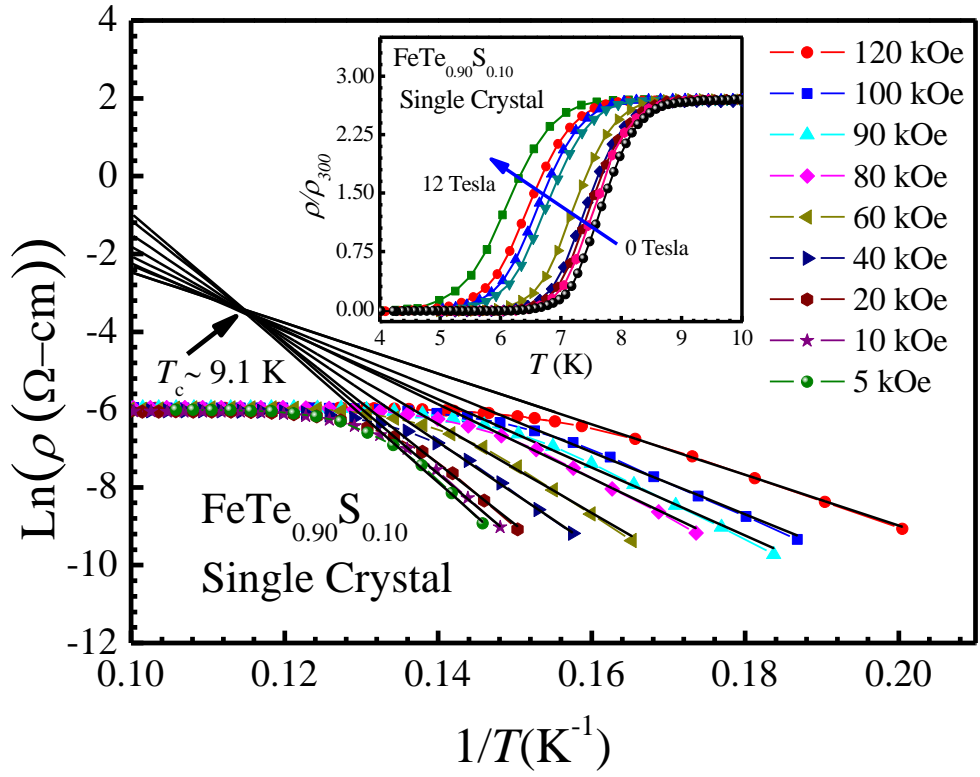

Fig 7(b)

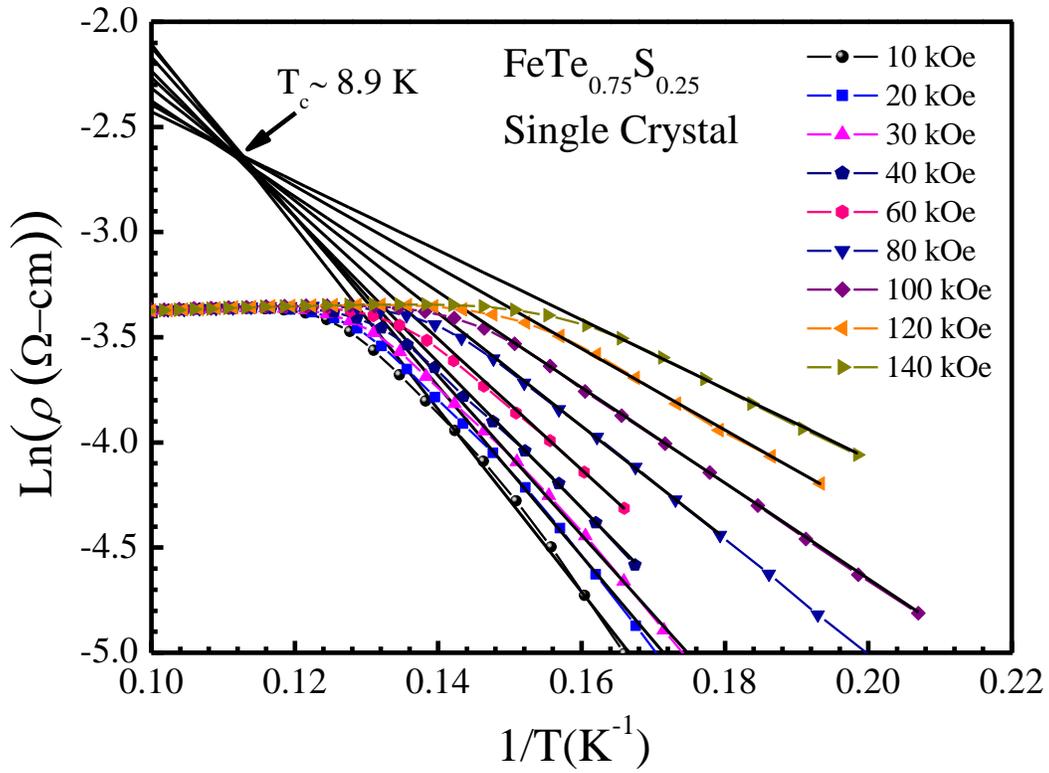

Fig 7(c)

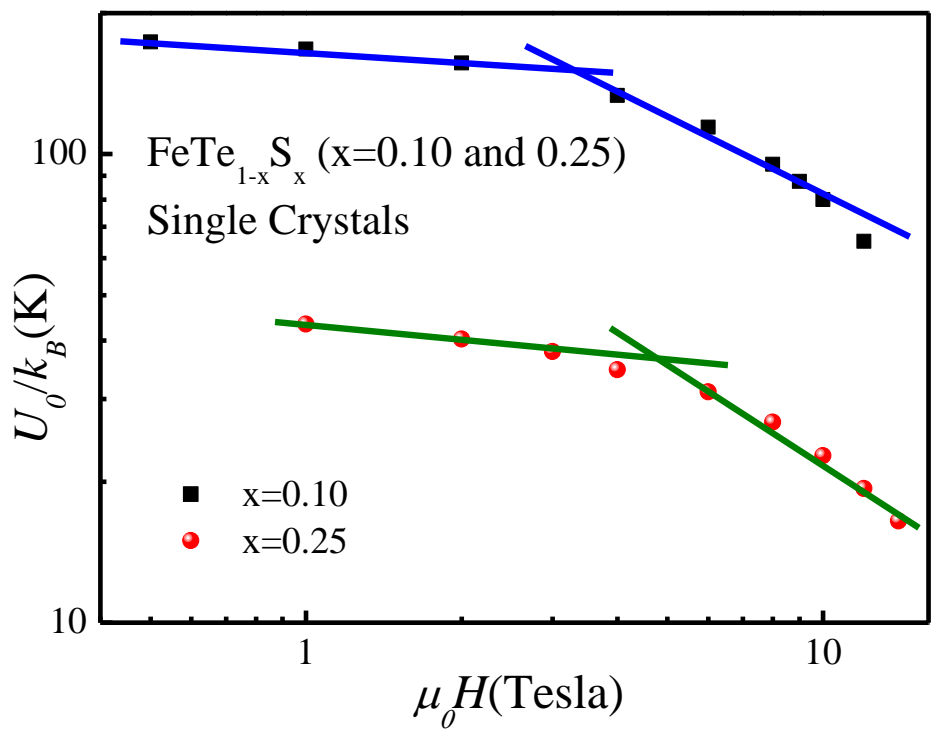

Fig 8

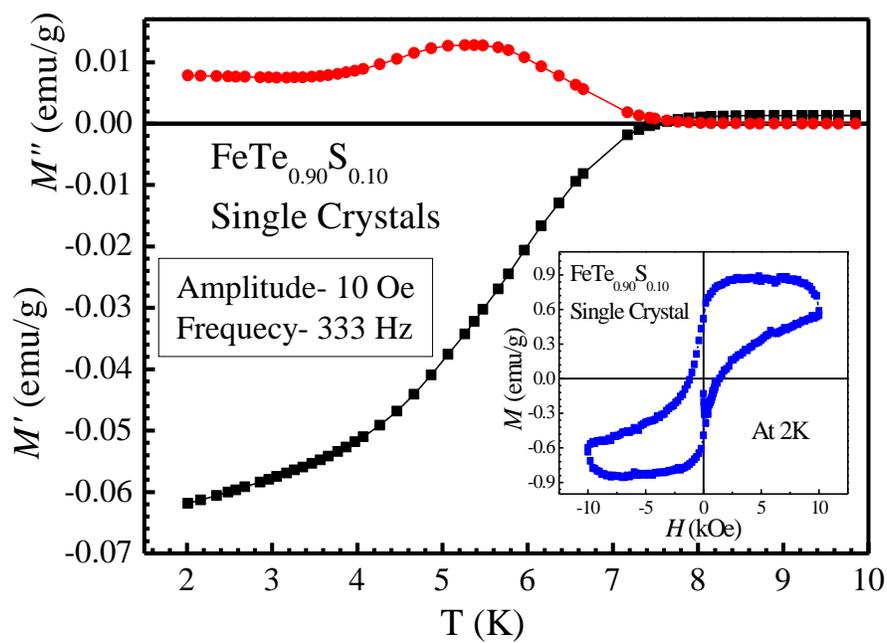

Fig 9

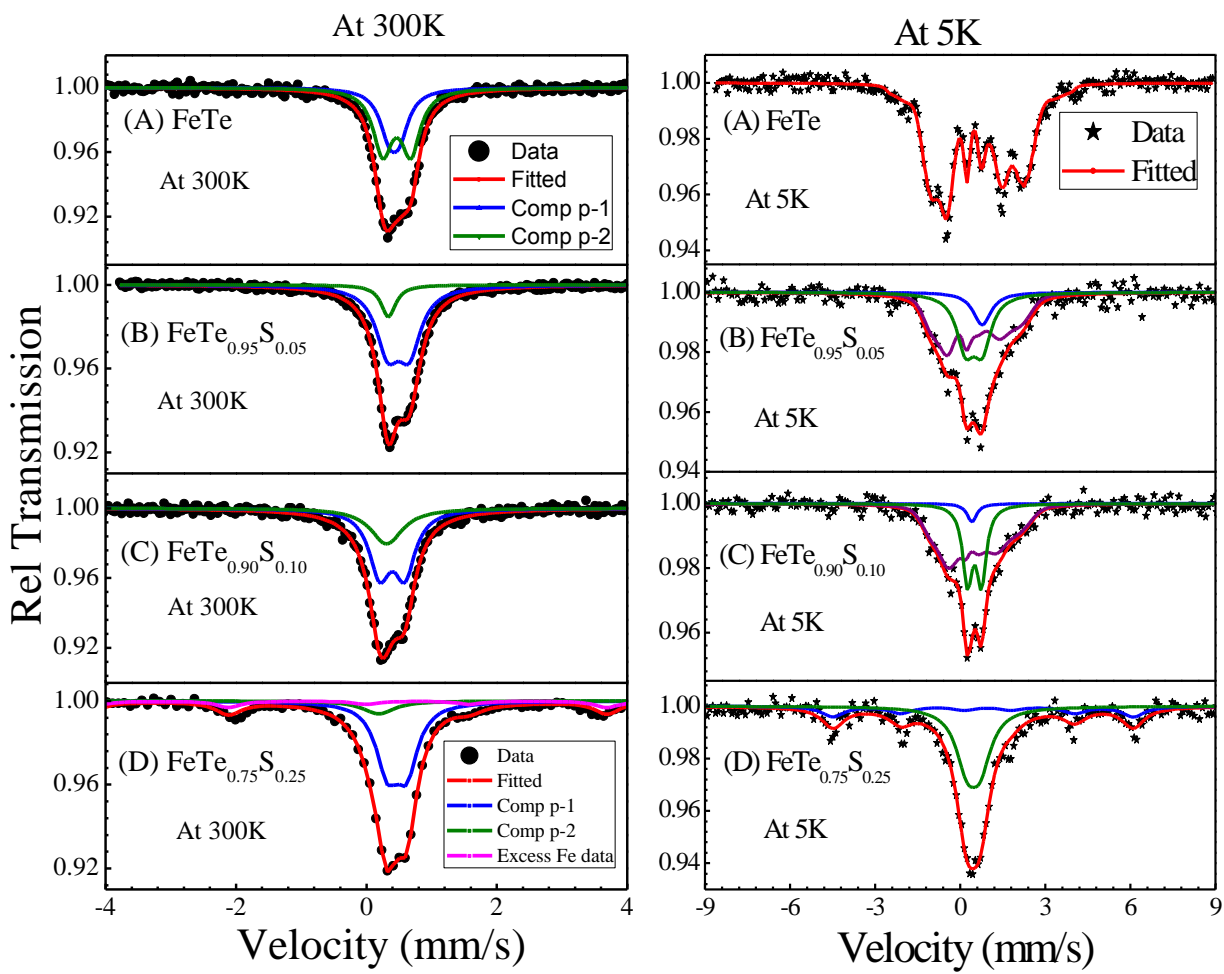